\documentclass{aa}
\usepackage{graphics}
\usepackage{epsfig}

\usepackage{txfonts}
\usepackage{color}
\def\red{\textcolor{black}}

\def\f1{f_{\rm I}}
\def\alphaD{\alpha_{\rm D}}

\def\mearth{M_\oplus}
\def\mjup{M_{\rm J}}
\def\msun{M_\odot}

\def\rhill{R_{\rm Hill}}

\def\aplanet{a_{\rm planet}}
\def\mplanet{M_{\rm planet}}
\def\mstar{M_{\rm star}}
\def\mdisk{M_{\rm disk}}
\def\Tdisk{T_{\rm disk}}

\def\fpg{f_{\rm D/G}}
\def\fpgsun{f_{\rm D/G, \odot}}
\def\alphaD{\alpha_{\rm D}}

\def\mearth{M_\oplus}

\def\aplanet{a_{\rm planet}}

\def\rhill{R_{\rm H}}

\def\msun{M_\odot}

\def\simgr{\,\hbox{\hbox{$ > $}\kern -0.8em \lower 1.0ex\hbox{$\sim$}}\,}
\def\simle{\,\hbox{\hbox{$ < $}\kern -0.8em \lower 1.0ex\hbox{$\sim$}}\,}
\def\beq{\begin{equation}}
\def\eeq{\end{equation}}

\def\simgr{\,\hbox{\hbox{$ > $}\kern -0.8em \lower 1.0ex\hbox{$\sim$}}\,}
\def\simle{\,\hbox{\hbox{$ < $}\kern -0.8em \lower 1.0ex\hbox{$\sim$}}\,}
\def\beq{\begin{equation}}
\def\eeq{\end{equation}}

\def\apj{ApJ}                 % Astrophysical Journal
\def\apjl{ApJ}                % Astrophysical Journal, Letters
\def\apjs{ApJS}               % Astrophysical Journal, Supplement
\def\ao{Appl.Optics}          % Applied Optics
\def\aap{A\&A}                % Astronomy and Astrophysics
\begin{document}

\title{\red{Extrasolar planet} population synthesis}
\subtitle{III. Formation of planets around stars of different masses}

\author{Y. Alibert\inst{1,2}, C. Mordasini\inst{\red{3,1}}, W. Benz\inst{1}}

\institute{Physikalisches Institut, University of Bern, Sidlerstrasse 5, CH-3012 Bern, Switzerland\\
 \email{[yann.alibert;wbenz]@space.unibe.ch}
\and
Institut UTINAM, CNRS-UMR 6213, Observatoire de Besan\c{c}on, BP 1615, 25010 Besan\c{c}on Cedex, France 
\and
Max-Planck Institute for Astronomy, K\red{\"o}nigstuhl 17, 69117 Heidelberg, Germany\\  
\email{mordasini@\red{mpia}.de} 
} 

%CHRIS
%\offprints{Y. Alibert \email{alibert@obs-besancon.fr}}

\date{Submitted/Accepted}
\abstract
{}
{We extend the models presented in Mordasini et al. (2009) to the formation of planets orbiting stars of different masses. We discuss the properties of the resulting synthetic planet population in term\red{s} of mass, orbit, and metallicity distributions. }
{The population synthesis calculations we use are based on the planet formation model developed in Alibert et al. (2005a), which self-consistently take\red{s} into account planetary growth and migration in an evolving  proto-planetary disk. Using this model, we generate synthetic populations of planets by following their growth in a large number of proto-planetary disks, whose properties (mass and lifetime) are selected in a Monte Carlo fashion using probability distributions derived from observations.}
{We show that the scaling of the proto-planetary disk mass with the mass of the central star has a direct and large influence on the properties of the resulting planet population. In particular,  the observed paucity of large mass planets orbiting 0.5 $\msun$ stars can directly be explained as resulting from a only slightly steeper than linear scaling. The observed lack of short period planet orbiting 2.0 $\msun$ stars can also be attributed to this scaling but only if associated with a decrease of \red{the} mean disk lifetime for stars more massive \red{than} 1.5 $\msun$. Finally, we show that the distribution of minimum mass and semi-major axis of our synthetic planets are statistically
comparable with observations.}
{}


\titlerunning{\red{Extrasolar planet population synthesis.} III}
\authorrunning{Y. Alibert et al.}

\maketitle

\section{Introduction}

Since the \red{pioneering} discovery of the first extrasolar planet orbiting a main sequence star by Mayor and Queloz (1995), more than 400 extrasolar planets have  been discovered, the majority of them by means of the radial velocity technique. After having essentially focussed on solar type stars, more and more surveys are targeting stars of different nature, in particular of different mass. From a theoretical point of view, these surveys are important since a global understanding of the processes at work during planet formation relies on the comparison of observations with model results in environment as varied as possible.

In the core-accretion scenario of planet formation, gas giant planets are believed to form in two steps. In a first step, the planet's core is formed by accumulation of planetesimals, in a way similar to the one generally accepted for the formation of terrestrial planets. When a critical mass is reached, the core starts to capture ever increasing quantities of nebular gas, leading to a phase of runaway gas accretion at the end of which the planet reaches its final mass. For a long time, this formation model suffered from a nagging timescale  problem. Unless the disk was quite massive, the formation time, as computed by the model, was found to significantly exceed observationally inferred proto-planetary disk lifetimes. In recent years, improving the physics handled by the models and/or considering additional physical processes has led to a significant speed-up of the formation process thereby removing this timescale problem. Indeed, theoretical models based on new opacity determinations of Podolak (2003) allowed formation of Jupiter-like planets in a few Myr (see e.g. Hubickyj et al. 2005). Moreover, extended core-accretion models taking into account migration of forming planets, as well as proto-planetary disk structure and evolution led to equally short formation time (see Alibert et al., 2004, 2005a). Finally, these latter models  have been shown to be compatible with our present day knowledge of JupiterÕs and Saturn's internal structure and atmospheric composition (Alibert et al. 2005b). 

Many different studies have already considered the formation of planets orbiting stars of different masses. Laughlin et al. (2004) have shown for the first time that the formation of gas giant planets orbiting low mass stars is inhibited. This is mainly due to the reduced Keplerian frequency in this environment\footnote{The Keplerian frequency $\Omega = \sqrt{\mstar G \over r^3}$, where $\mstar$ is the mass of the primary, \textit{\red{G}} the gravitational constant, and \textit{\red{r}} the distance from the central star, gives the typical timescale of collisional processes involved in the formation of planets.}, and to their assumption of a strong scaling between disk and primary mass. Ida and Lin (2005) used their planet formation model  to study the influence of the central star mass on the characteristics of the planets. To relate the mass of the proto-planetary disk to the one of the primary star, they used a scaling relation of the type: $\mdisk \propto \mstar ^{\alpha_D}$, where $\alpha_D$ was allowed to vary between 0 and 2. Their study shows that close-in Neptune-mass ice-giant planets should be common around low mass stars, while Jupiter-mass gas-giant planets should be quite rare. On the opposite, the high-mass peak of the planetary mass function grows in importance around high mass stars. 

Finally, Kornet et al. (2006) have calculated the influence of the central star mass on the evolution of the protoplanetary disk, 
by focusing in particular on the redistribution of solids in the disk. Using a simple model to relate the resulting structure of the proto-planetary disk and the probability of presence of a giant planet, they have concluded that, globally, the probability to harbour a Jupiter mass planet increases with a decreasing primary mass. Interestingly enough, their conclusions are very different from the ones by Laughlin et al. (2004) and Ida \& Lin (2005). Indeed, in these studies the surface density of planetesimals increases  with the mass of the primary, whereas in Kornet et al. (2006) it decreases due to the fact that the redistribution of solids is more efficient in low mass star environments. Present day observations suggest that the probability to harbour an observable giant planet increases with the mass of the central star, which shows that the redistribution of solids in protoplanetary disks is not the dominant process in the early stages of planet formation.

In a recent paper, Mordasini et al. (2009a, paper I) have presented planet population synthesis calculations based on the extended core-accretion formation model of Alibert et al. (2005a) which takes into account migration and disk evolution. This model has been shown to reproduce some of the bulk properties of the giant planets in our own Solar System (Alibert et al. 2005b), and of the three \red{Neptune} mass planet system discovered around HD69830 (Lovis et al. 2006, Alibert et al. 2006). Population synthesis calculations rely on the calculation of the formation of thousands of planets, each one assuming a different set of initial conditions. In particular, the mass of the proto-planetary disk, its lifetime, its metallicity (more exactly the planetesimals-to-gas ratio in the disk), and the starting location of the planet's initial embryo have to be specified. The distribution functions for these initial conditions are taken, as far as possible, directly from astronomical observations. Furthermore, by taking into account the detection  bias introduced by radial velocity surveys (Naef, 2004), it was possible to compare in a statistical way the observed population of giant planets with the  synthetic ones. Using Kolmogorov-Smirnov (KS) tests, Mordasini et al. (2009b, paper II) showed that this approach yields a planet population whose mass and semi-major axis distributions match the properties of the actually observed planets to a confidence level of close to 90 \%.

In this paper, we extend the calculations presented in papers I and II to the formation of planets orbiting stars of different masses. In order to isolate the effects of varying the mass of the star, we use the same formation model varying only the central star mass. Note that, since we also consider models for which the properties of the proto-stellar disk vary with the mass of the primary, this also leads for these cases in changed disk characteristics. 

Since we not only use the exact same procedure as explained in detail in paper I and II but also all the same value for all numerical parameters, we will not discuss our approach in any detail here but refer the reader to these papers. The present paper is organized as follows: in Sect. 2, \red{we} discuss the influence of the mass of the primary on the formation process itself as well as on the corresponding initial conditions. In Sect. 3, the outcome of model in which the mass of the central star is varied from 0.5 $\msun$ to 2.0 $\msun$ are presented. In particular, we discuss the changes in the properties of the resulting planets (mass and semi-major axis, composition) as well as the influence of metallicity in the formation process. Finally, Sect. 4 will be devoted to the discussion and conclusions.

\section{Formation model and initial conditions} 

\subsection{Influence of the central star mass}

The planet formation model we use is a simplified version of the one presented in Alibert et al. (2005a). As we have already described this model extensively in paper I,  we will not repeat it here. Our approach consists in following the growth of a single planetary embryo initially located at a given distance of the central star in a disk consisting of gas and solids. { Note that the opacity is not reduced compared to interstellar values
(e.g. Pollack et al. 1996, Hubickyj et al. 2005). In our calculation, the effect of such a reduction is indeed rather small. The reason for that is that thanks to migration (even with the strongly reduced type I migration rate), cores never get completely cut from a supply of fresh planetesimal to accrete, without the need of having  the envelope growing (where the opacity matters) in order to expand the solid feeding zone, as it is the case for a strict in situ formation. 
These effects are discussed in Mordasini et al. (2010).}

The actual mass of the central star  enters  as a parameter in many of the physical processes involved during planet formation. Explicitly, it enters in: 
\begin{enumerate}
\item The Hill's radius ($\rhill = \aplanet \left( { \mplanet \over 3 \mstar }\right)^{1/3}$, where $\mplanet$ and $\mstar$ are the planet and star masses, and $\aplanet$ is the planet's semi-major axis). This radius is a measure of the size of the planet's feeding zone, and (indirectly) of its envelope size and hence mass. Since it inversely scales with stellar mass, planets forming in orbit of a more massive stars can accrete planetesimals originating from a smaller region of the proto-planetary disk.
\item The type I migration rate. Type I migration rate, which is relevant for low mass planets, is reduced for more massive central stars (see Tanaka et al. 2002).
\item The disk structure. On one hand, increasing the mass of the central star results in a stronger vertical component of the gravity in the disk. On the other hand, because the viscous dissipation rate is a function of Keplerian frequency, increasing the stellar mass leads to hotter disks. Numerical calculations show that the first effect dominates over the second and that disks orbiting more massive stars, everything else being equal, are generally thinner.
\item  The gravitational radius $R_{\rm g}$ (see Veras and Armitage, 2004), which sets the region in the disk where gas can be photo-evaporated. This parameter depends linearly upon the central star mass (see Adams et al. 2004).
\item The Keplerian frequency which, among other things, governs, the accretion rate of solids.
\end{enumerate}

In addition, there are stellar mass dependencies in the computation of the initial conditions. Initially, all our disks are characterized by a power law surface density of index $-3/2$. However, the temperature and pressure structures of these disks are functions of the mass of the central star. Hence, the stellar mass determines the location of the iceline where the surface density of solids jumps by a factor of about four.  All other parameters being equal, the iceline moves outwards with increasing stellar mass by 1 to 2 AU (depending on the disk's mass) for a stellar mass varying from $\mstar = 0.5 \msun$ to $\mstar = 2.0 \msun$. An analytical fit using our nominal  value for the viscosity parameter $\alpha  = 7 \times 10^{-3}$ (Shakura \& Sunyaev, 1973), provides the following approximation for the initial position of the iceline 
\begin{equation}
 {r_{\rm ice} \over {\rm AU} } = \left( \Sigma_{\rm 5.2 AU} \over 10 {\rm g/cm^2} \right)^{0.44} \times \left( { \mstar \over \msun } \right)^{0.1}
\label{eq_iceline}
\end{equation}
where $ \Sigma_{\rm 5.2 AU}$ is the initial gas surface density at 5.2 AU\footnote{in our model, the location of the iceline is set by the initial conditions and does not vary with the evolution of the disk, see paper I}. 

{ We recall here that we only compute the disk structure in the inner 30 AU of the protoplanetary disk. This choice is governed by the fact that all
planets, in these calculations, form in the central parts of the disk. In order to solve the diffusion equation, we have to assume a boundary condition at 30 AU.
In the calculations presented here, we assume , as in Alibert et al. (2005), that the mass flux at 30 AU is null. In this sense, we overestimate the gaseous 
mass in the inner 30 AU, since part of the gas in reality goes beyond 30 AU for angular momentum conservation. However, the viscous timescale at 
30 AU is long enough so that the precise boundary condition there has a low influence on the planet forming region (below 10 AU or so). In addition, note that 
the lifetime of disk is, by construction, similar to the  observed one, at least for solar type stars. Indeed, the photoevaporation rate is adjusted so that 
our disk population has a disk lifetime probability consistent with observations. Note that changes in the disk lifetime only result from variations in the
assumed photoevaporation rate and {\it not} from variations in the assumed viscosity parameter.

Finally, }as in Alibert et al. (2005), the structure of the disk is calculated without taking into account the irradiation of the central star, which itself would depend on the mass of the primary. Since higher mass stars tend to have higher luminosities, this  would also translates into an iceline located at larger distances.

\begin{figure}
\resizebox{\hsize}{!}{
\includegraphics[angle=0]{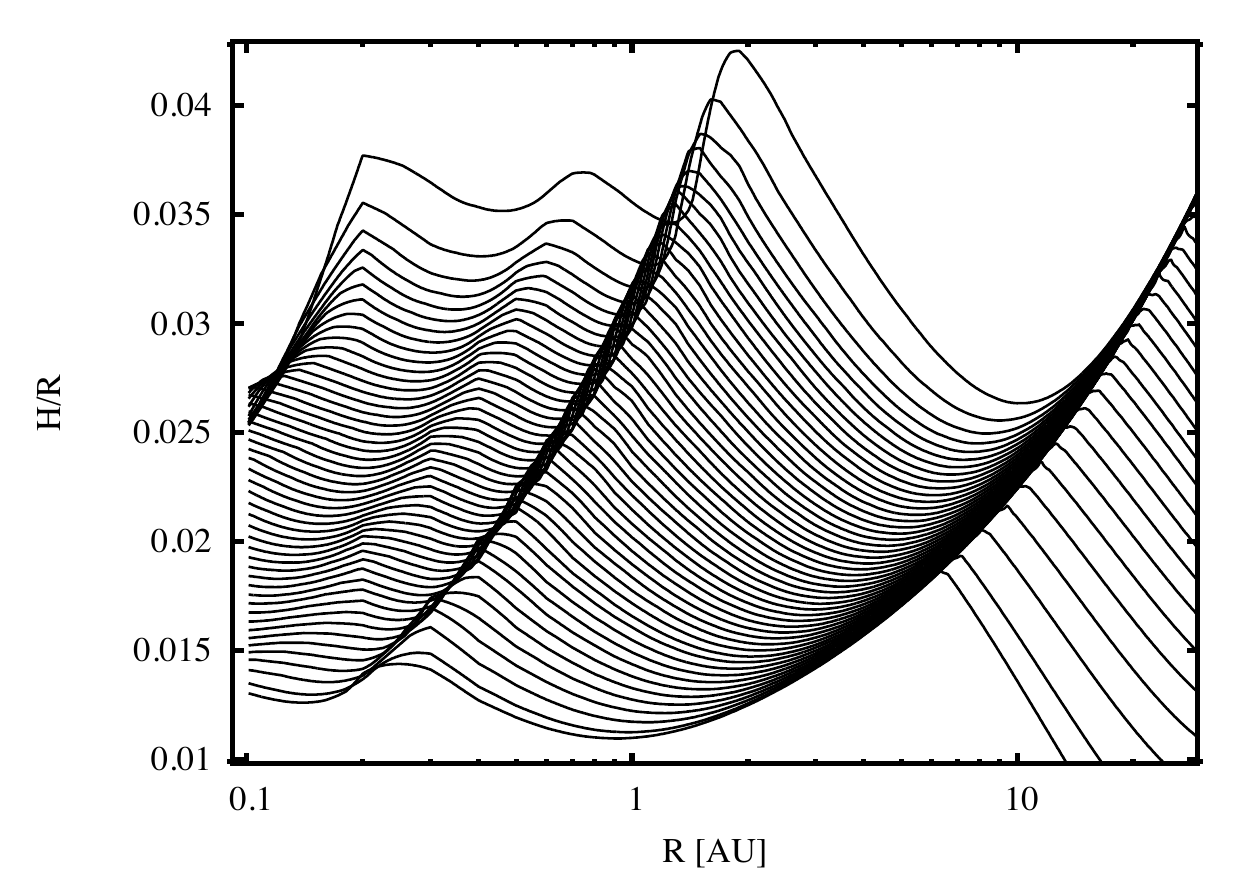}}
\caption {Disk aspect ratio (thickness over distance to the star) as a function of distance to the central star. Each line represents a different epoch, from 0.1 Myr *uppermost line) to 5.2 Myr, with a 0.1 Myr step. The initial mass
of the disk is 0.016 $\msun$, and the lifetime is 5.3 Myr. The mass of the central star is 1.0 $\msun$.} 
\label{Hdisk}
\end{figure}

{In order to ease the comparison with other disk models, Fig. \ref{Hdisk} presents the evolution of the disk aspect ratio (which is important for the switch between type I and type II migration) for a "typical" disk, namely
a disk with a mass of 0.016 $\msun$, a lifetime of 5.3 Myr, around a 1.0 $\msun$ star. }

Finally, the time required to grow the seed embryo used in our calculations depends on the Keplerian frequency as well as on the actual distribution of solids. This time determines the time offset between disk evolution and the start of embryo growth and hence the time available to accrete solids and gas before the disk disappears (see paper I). For a fixed surface density, numerical calculations show that the embryo formation timescale at a fixed distance decreases within increasing stellar mass. This effect is further amplified by the fact that disks orbiting high mass stars are more massive than their counterparts orbiting low mass stars. We note that depending upon the location of the iceline, disks orbiting smaller mass stars may have a locally higher surface density than disks orbiting more massive stars, but this only concerns a small number of disks and is limited to small regions of these disks.  

\subsection{Parameters and initial conditions}

In this paper, we aim at studying in a global framework the formation process of planets orbiting stars of different mass. As a consequence, we will use for the parameters of the model the same values as in paper I. Depending upon the scaling of the disk with the mass of the primary, the disk mass and lifetime will represent an exception to this rule. In addition, since part of our work focusses on the frequency of close-in planets predicted to orbit stars of various masses, we now follow the planetary migration and growth until the planet crosses the inner boundary of the protoplanetary disk located at 0.1 AU. This means that the inner edge of the planet feeding zone, which was located at $\aplanet - 4 \rhill$ in paper I and II, is now located at the greater of this former value, and 0.1 AU. This modification has no effect on the mass growth of planets.

The two key parameters characterizing the formation model are the $\alpha$ parameter entering in the determination of the viscosity and the type I migration rate parameter $f_I$. As shown in paper II, population synthesis calculations allow to reproduce the semi-major axis and mass distributions for single giant planets orbiting G stars using $\alpha = 7 \times 10^{-3}$, and $f_I = 0.001$. Other parameters of the model are taken from the nominal model in paper I, and are summarized in table \ref{parameter_model}, where  $f_{\rm D/G,\odot}$ is the assumed planetesimals to gas ratio in the planet forming region for a solar metallicity system.  We stress that, for simplicity, we do not take into account the dependance of star's \red{metallicity} with respect to  mass. Even if M dwarfs are known to be slightly metal poor, this effect is not included in our calculations. We justify this by noticing the much larger uncertainty existing on the relationship between disk mass and primary mass.

\begin{table} \label{tab:variedparameters}\begin{center}
 \caption{Parameters for the simulations used in the paper (except if \red{explicitly} mentioned).}
\begin{tabular}{ll}
\hline
Feature & Values \\
\hline
Type I migration reduction factor $\f1$  & 0.001 \\
Viscous parameter $\alpha$ & $7 \times 10^{-3}$  \\
Initial exponent gas disk $\Sigma(a,t=0)$ & -3/2\\
Rockline included & no \\
Iceline included & yes\\
Outer radius of the computational disk & 30 \\
Inner radius of the computational disk & 0.1 \\
$f_{\rm D/G,\odot}$ & 0.04\\
\end{tabular}
\label{parameter_model}
\end{center}
\end{table}

The determination of the relation between the disk mass and the central star mass \red{has} been the subject of numerous observational studies.  In our model, we assume that the two quantities are linked by a relation  of the type $\mdisk \propto \mstar^{\alphaD} $. Based on H$\alpha$ line profiles, Muzerolle et al. (2003), and Natta et al. (2004) have inferred that the disk accretion rate roughly scales with the square of the primary mass. Using our disk model we have determined the dependance of this accretion rate at the inner edge of our computational disk on the value of $\alphaD$. Our calculations show that $\alphaD \sim 1.2$ allows to roughly reproduce the data from Muzerolle et al. (2003) and Natta et al. (2004). We set this value as our nominal value but also consider the cases for which $\alphaD= 0$ and $2$ in an effort to provide bracketing results. Finally, in order to ensure that the disks considered are dynamically stable, we  only consider disks whose mass is lower than a tenth of the central star mass regardless of the value of $\alphaD$.

The observational determination of the disk lifetime orbiting stars of different mass could only be undertaken in recent years thanks to a growing number of observations. By correlating the fraction of stars showing an infrared excess with their mass for different stellar clusters, Kennedy and Kenyon (2009) were able to show that disks orbiting stars more massive than $\sim 1.5 \msun$ have shorter lifetimes. While a reliable and accurate dependency of disk lifetime is still difficult to extract from the current set of observations,  the data seem to be \red{consistent} with disk lifetimes decreasing as $\mstar^{-1/2}$, for $\mstar > 1.5 \msun$, and disk lifetimes \red{independent} of the central star mass for less massive stars. In the following, we adopt this dependency but consider also a case in which the disk lifetime is independent of the stellar mass for all the masses considered (see section \ref{hotplanets}).

\section{Results}
\label{results}

\subsection{Minimum mass \textit{versus} semi-major axis diagrams}
\label{am}

\begin{figure}
\resizebox{\hsize}{!}{
\includegraphics[angle=0]{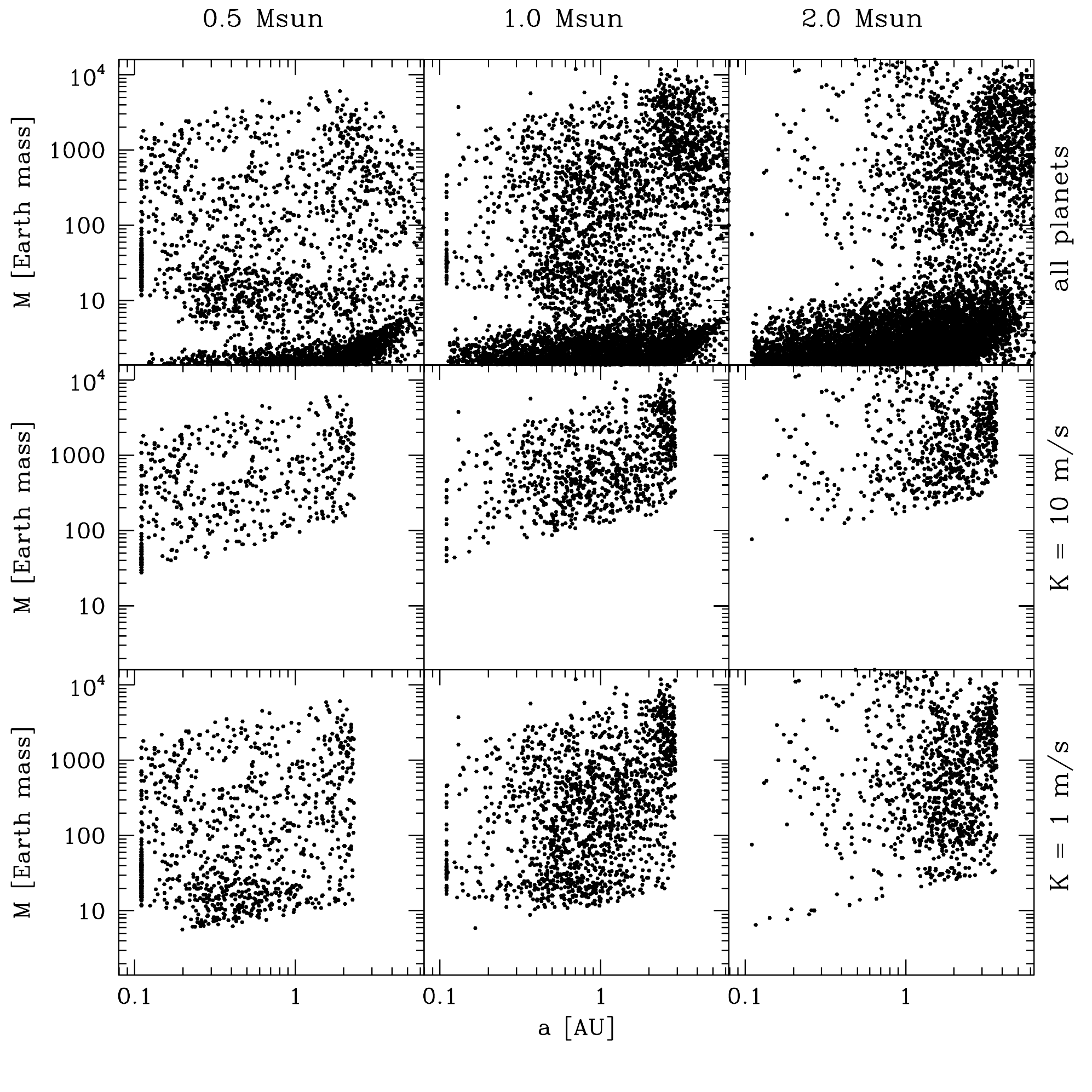}}
\caption {\textbf{Upper panels}: mass \textit{versus} semi-major axis diagrams for the nominal model. The mass of the primary is equal to 0.5 $\msun$ (left panel), 1.0 $\msun$ (central panel), and 2.0 $\msun$ (right panel). 
\textbf{Middle panels}: mass \textit{versus} semi-major axis diagrams for the nominal model, as can be detected with a 10 m/s accurate spectrograph, and with a period lower than 5 years. The masses of the primary are the same as in upper panels\red{.} For each planet, the inclination angle between the line of sight and the orbital plane is chosen at random, in order to compute the resulting Doppler shift.
\textbf{Lower panels}: as middle panels, but for a 1 m/s accurate spectrograph,
and a period lower than 5 years. Note that in all the diagrams, planets
crossing the inner boundary of the calculated disk are assumed to stay at 0.1 AU. The number of synthetic planets is equal to 30000 for each primary mass.} 
\label{a_m_nommodel}
\end{figure}

\begin{figure}
\resizebox{\hsize}{!}{
\includegraphics[angle=0]{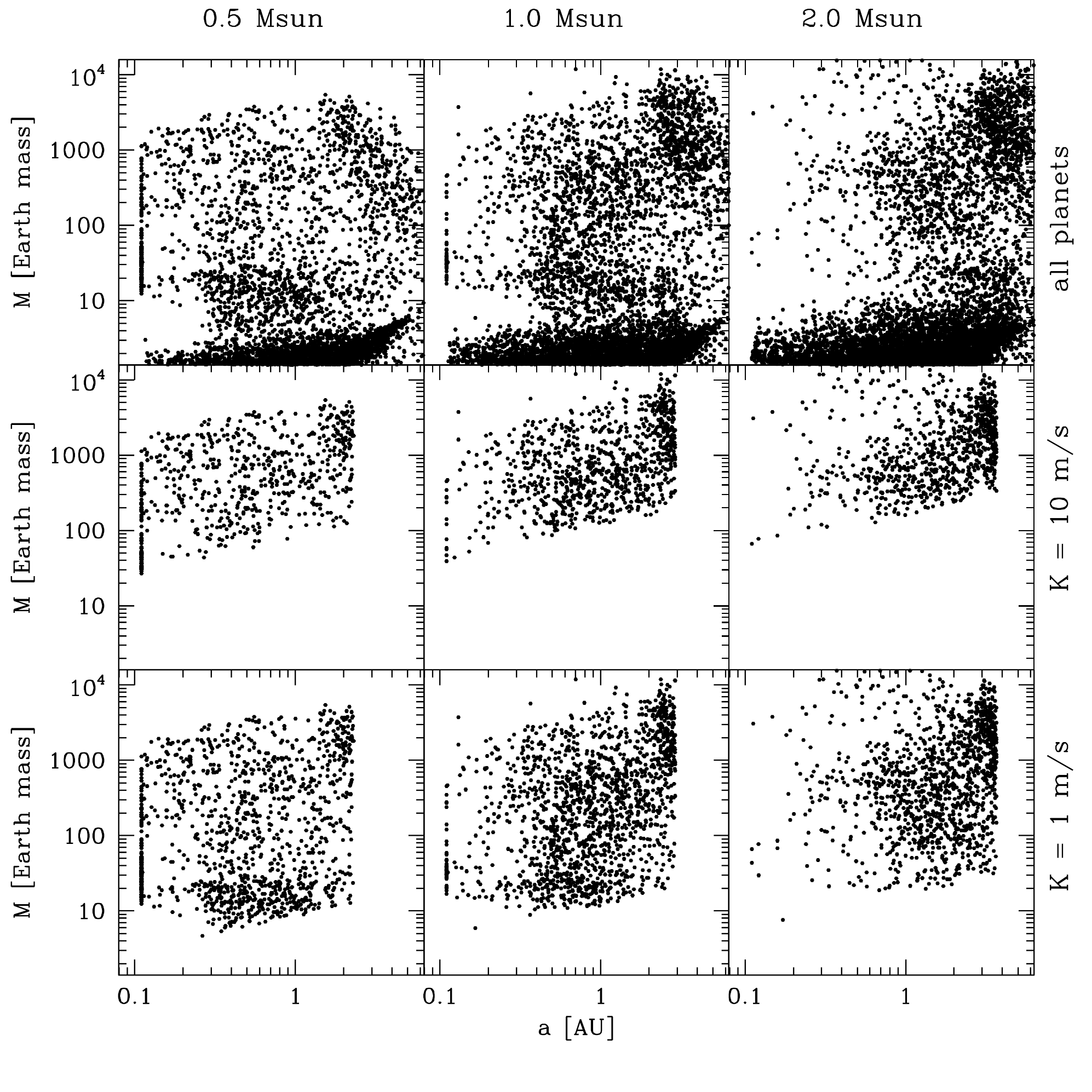}}
\caption {Same as Fig. \ref{a_m_nommodel}, but for a disk mass \red{independent} of the primary mass ($\alpha_{\rm D}=0$).} 
\label{a_m_alphaD0}
\end{figure}

\begin{figure}
\resizebox{\hsize}{!}{
\includegraphics[angle=0]{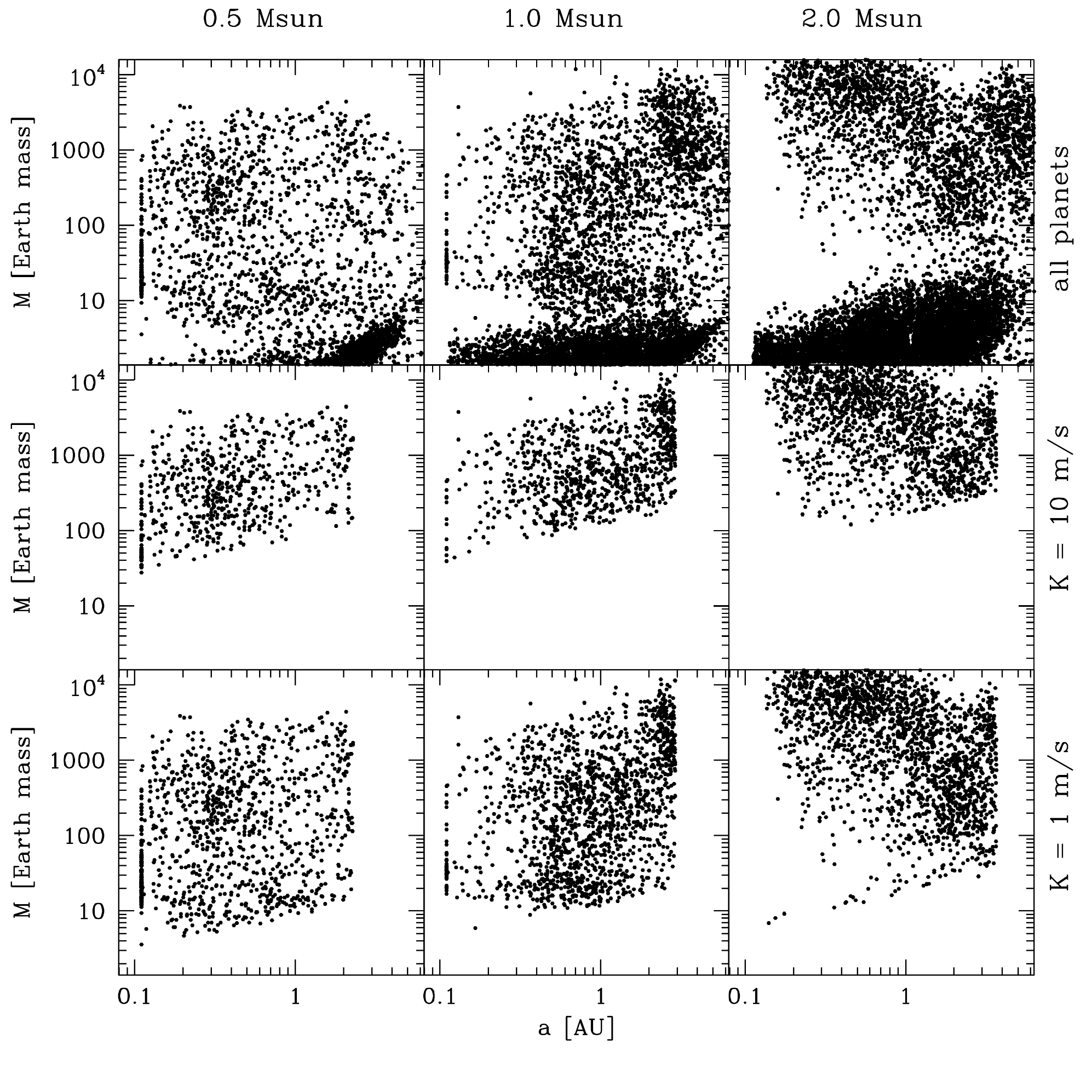}}
\caption {Same as Fig. \ref{a_m_nommodel}, but for $\alpha_{\rm D}=2$ (disk mass strongly depending on the primary mass).} 
\label{a_m_alphaD2}
\end{figure}

Fig. \ref{a_m_nommodel} to \ref{a_m_alphaD2} show the predicted mass \textit{versus} semi-major axis distributions of synthetic planet populations computed using either our  nominal model ($\alphaD = 1.2$), a model with a disk mass \red{independent} of the primary mass ($\alphaD = 0$), and a  model with a disk mass strongly dependent upon it ($\alphaD = 2$). In each of these figures, the columns from left to right correspond to central stars with masses to 0.5, 1, and 2.0 $\msun$ respectively. To allow for a comparison with observations, the middle and lower panels in each figures display the sub-population  whose induced Doppler semi-amplitude is larger than 10 m/s or 1 m/s, and orbital period less than 5 years\footnote{Later, these planets will abusively be named "detectable with a 10 m/s or 1 m/s accuracy spectrograph".}.

\subsection{Theoretical mass function}

As planets migrate inwards during their formation, they might cross the inner boundary of our computational disk. When this happens, we freeze both the mass and the position of the planet. In our model, the final characteristics of these planets are therefore ill-defined as they might be accreted by the central star, or might survive provided a stopping mechanism is at work in the inner parts of the disk. These planets might also grow further in mass due to accretion of gas flowing from the outer parts of the disk toward the inner regions. We do not take these effects into account and simply fix the position of these planets at 0.1 AU and keep the mass they had at the time they crossed this limit.

\begin{figure}
\resizebox{\hsize}{!}{
\includegraphics[angle=0]{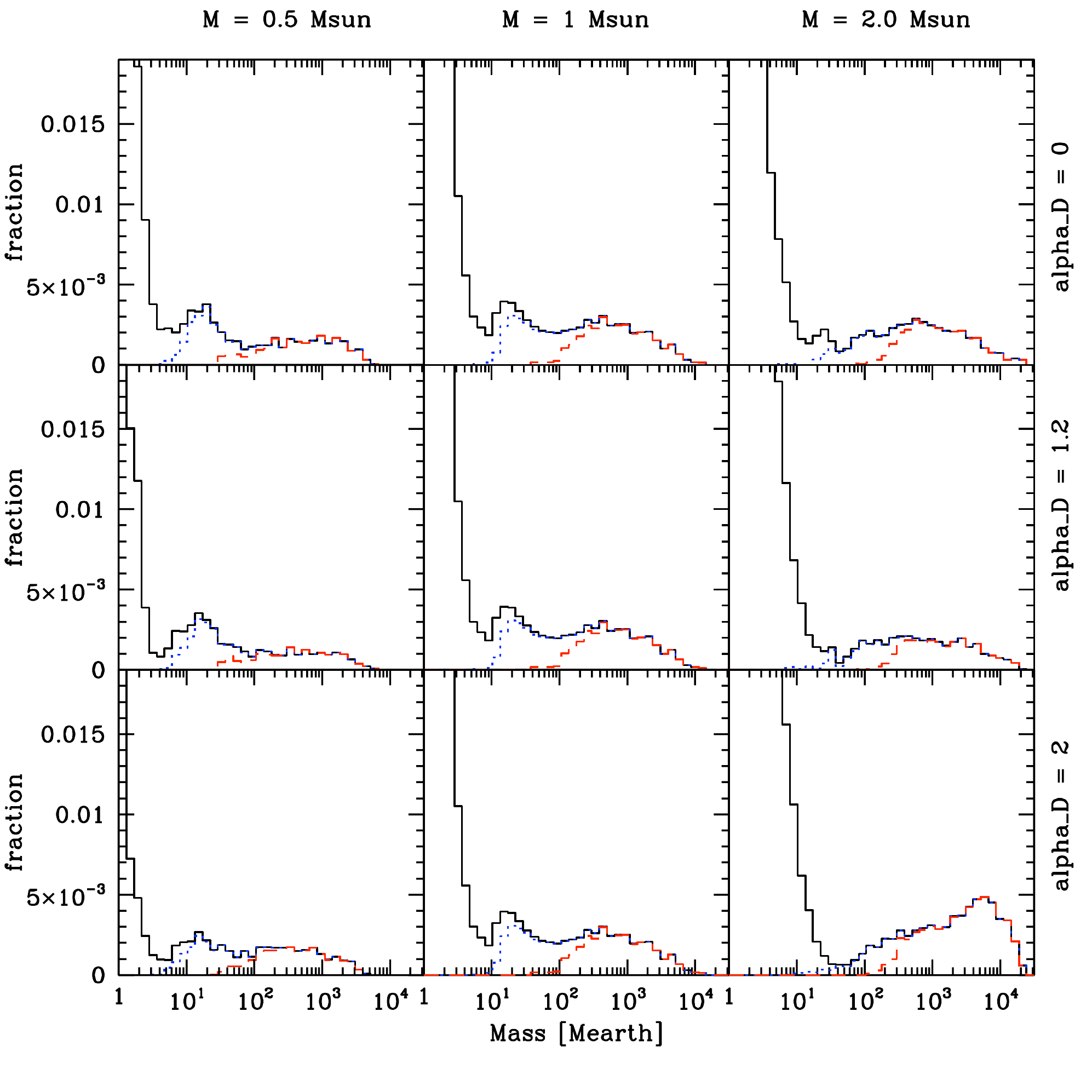}}
\caption {Planetary Initial Mass Function for different values of the primary mass (given at the top of each column) and different values of  $\alphaD$  (indicated on the right). In each panel, 30000 formation models have been considered. The y axis presents the fraction of resulting planets in a given mass bin. The solid black line gives the mass function of all planets, whereas the red dashed and blue dotted lines correspond to planets observable assuming respectively a 10 m/s and 1 m/s accuracy spectrograph.} 
\label{histomass}
\end{figure}

Fig. \ref{histomass} gives the planetary initial mass function for three values of the primary mass (0.5, 1.0 and 2.0 $\msun$) \red{and} different values of $\alpha_{\rm D}$. Three lines are displayed in each panel, corresponding to the full theoretical population and the planets that can be detected with a 10 m/s and a 1m/s accuracy spectrograph (see caption). \red{The} planets can be separated in three different classes (corresponding to the three peaks clearly seen for $\mstar = 0.5 \msun$). The first class of planets are low mass planets (a few Earth masses). They correspond to planets having accreted all the planetesimals present in their feeding zone and have subsequently never grown massive enough to \red{start} type II migration. Because in the nominal model type I migration is strongly reduced, they remain close to their initial location and their growth is being quenched by exhaustion of the reservoir of planetesimals to accrete. Their final mass is therefore roughly given by the isolation mass at the initial location of the embryo. 
The second class of objects are Neptune-mass planets. These planets are formed in more massive disks, which allow them to become massive enough to start type II migration, thus increasing further their effective feeding zone. However, their mass never reaches the critical mass, hence they never enter runaway gas accretion and remain of Neptune-like structure. Finally, the third case correspond to planets equal or more massive than Jupiter, which form in quite massive disks and start their formation at large distance. These planets are able to reach the critical mass and enter a runaway gas accretion phase. The formation tracks of these planets (not shown here) are qualitatively very similar to the ones described in paper I, to which the reader is \red{referred} to for more details.

Our results show that the form of the mass function is similar for every value of $\alphaD$ and primary mass. However, some differences can be seen. The most numerous planets are in the \red{S}uper-\red{E}arth regime (below $10  \mearth$), and their fraction does not strongly depend on the mass of the star, or the assumed $\alphaD$ value. However, for $\alphaD = 2$ around  $2.0 \msun$ stars, \red{Super-Earth} planets are slightly more massive (see the thickness of the first peak, which increases). Indeed, these \red{planets} \red{essentially} accrete all the mass in their feeding zone, reaching the local isolation mass. The isolation mass depends on the local surface density of solids (which increases for  $\alphaD = 2$ around $2.0 \msun$ stars), and on the size of the feeding zone (which decreases for $2.0 \msun$ stars, due to the increased gravity of the star). Therefore, planets are more massive only when the effect of the surface density is large enough to counteract the effect of the central star mass, namely for $\alphaD = 2$ and $2.0 \msun$ stars.

For higher mass planets, one can notice that the histograms for Neptune mass planets and massive planets evolves in an opposite way. Neptune mass planets are more numerous around low mass stars compared to high mass stars. In particular, it is interesting to note that very few Neptune mass planets are predicted around 2.0 $\msun$ stars. This results essentially from the evolution of the disk mass (the effect is more pronounced for $\alphaD = 2$, see the maximum mass of planets around $2.0 \msun$), and can be explained as follows: when disks are more massive, planets are more likely to accrete enough mass and to reach the critical mass and enter in the "massive planets" population. 

Note that, as the primary mass increases, the size of the planetÕs feeding zone decreases (all the other parameters being kept constant), and one could conclude that this should result, on average, in lower mass planets. However, this effect is clearly offset by the increase in disk mass and the fact that, since many of the processes time scales vary with the Keplerian frequency, massive embryos form earlier around massive stars. All these results are qualitatively in agreement with Ida \& Lin (2005) and Laughlin et al.  (2004). While the formation of giant planets orbiting M dwarfs is indeed less probable,  it is important to stress that such planets can form around M dwarfs within the core accretion mechanism. 

Interestingly enough, for primary stars ranging from 0.1 to 2.0 $\msun$ the mass function for planets more massive than 100 $\mearth$ can be approximated by the following simple analytical law:
$$ \mplanet = M_{\rm planet, 1 \msun} \times \left( {\mstar \over \msun} \right)^\gamma$$
where $M_{\rm planet, 1 \msun}$ is the mass function around Solar mass stars, and $\gamma = 0.9$ provides good results for the nominal case ($\alphaD = 1.2$) although the match is somewhat worse for stars below 0.3 $\msun$. For $\alphaD = 0$, $\gamma = 0.8$ provides a reasonable match, whereas for $\alphaD = 2$, larger values of $\gamma$ are required (around 1.3), but the match is significantly worse in particular for high mass stars.

\subsection{Close-in planets}
\label{hotplanets}

Massive planets orbiting at close distances to their stars (hot jupiters) are the easiest ones to detect by radial velocity surveys.  So far, they represent about $10-20 \%$  of all the detected planets orbiting stars less massive than $\sim 1.5 \msun$. Interestingly enough, for more massive stars jupiter-like planets are only discovered at significantly larger distances (greater than $\sim 0.5$ AU; see e.g. Wright et al. 2009). Since these planets are the easiest ones to detect, this lack of hot jupiters orbiting massive stars is really \red{striking} and cannot be accounted for by some observational bias. 

\begin{figure}
\resizebox{\hsize}{!}{
\includegraphics[angle=0]{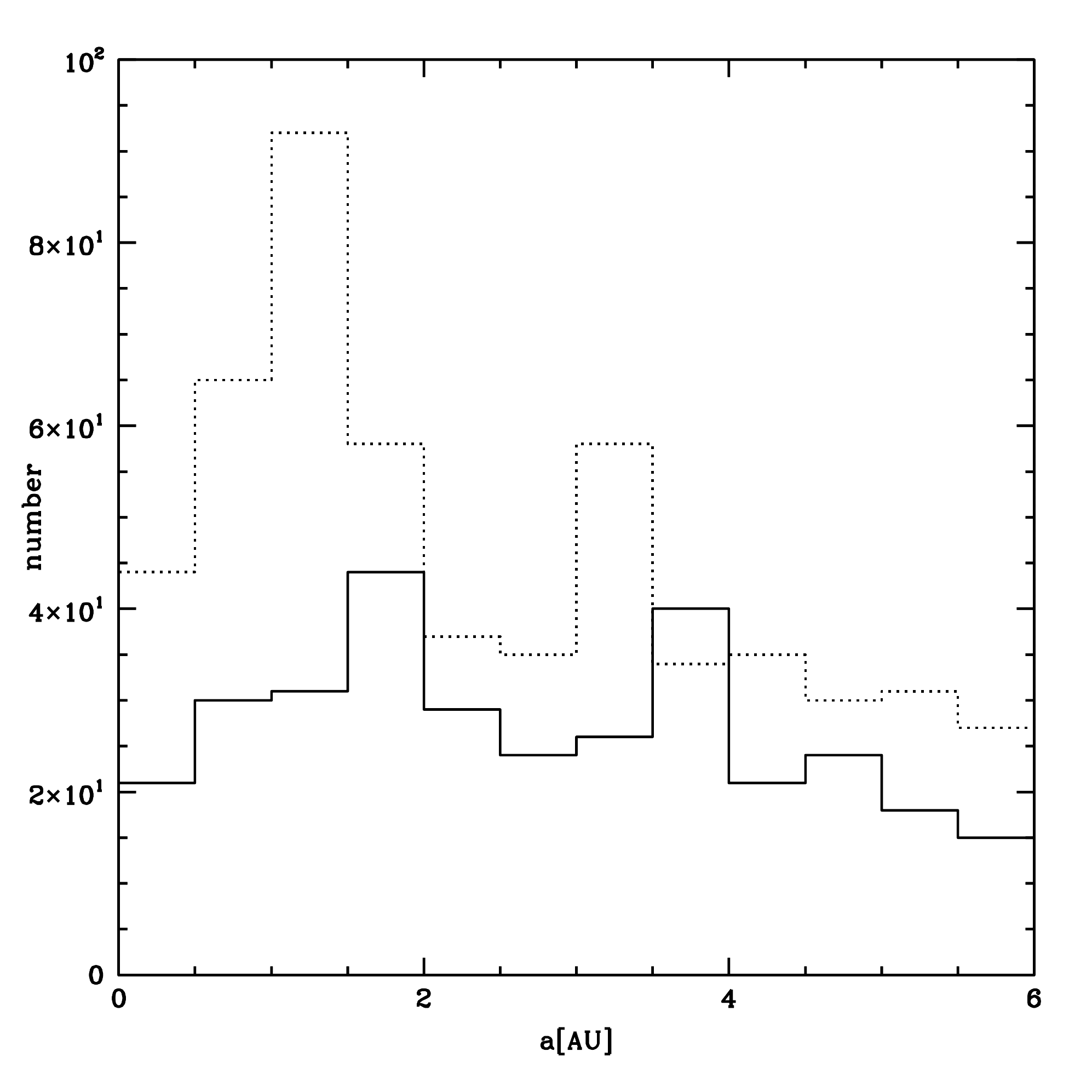}}
\caption {Histogram of final semi-major axis for detectable synthetic planets orbiting 2.0 $\msun$ stars, assuming a 10 m/s accuracy spectrograph. Dotted line: disk lifetime are similar as for disks orbinting solar type stars, solid line: the disk lifetime is 30\% shorter.
In each case, 9000 formation models have been calculated. The number of observable planets is 424 and 621 respectively for
the solid and dotted lines.} 
\label{effetTdisk}
\end{figure}

It has been proposed in the literature that this lack of massive planets orbiting massive stars at close distances could be due to the shorter lifetimes of disks orbiting massive stars (see e.g. Currie 2009). In order to test this hypothesis, we have re-calculated the theoretical planet population orbiting 2.0 $\msun$ stars but assuming that the distribution of disk lifetime is identical to the one orbiting solar type stars. Fig. \ref{effetTdisk} shows the comparison between the two resulting populations. We confirm the fact already mentioned in the literature (e.g. Johnson et al. 2007) that assuming identical disk lifetimes results in the formation of a large number of hot planets orbiting 2.0 $\msun$ stars which would have been easily detected. Conversely, reducing the liftetime of disks orbiting 2.0 $\msun$ stars by 30 \% shorter (following the  $\Tdisk \propto \mstar^{-1/2}$ suggested by Kennedy \& Kenyon 2009 for high mass stars) results in a planet population without hot planets and only very few orbiting below 0.5 AU. 

This effect can be explained in terms of a reduced overall migration for reduced disk lifetimes. Indeed, planets in orbit about massive stars grow larger in mass and more rapidly (see previous section) while at the same time the disk is evolving more rapidly. These two effects combine to hasten the transition from type II disk-dominated migration to type II planet-dominated migration which is the migration regime characterized by a ratio of planet mass to local disk mass exceeding unity. Compared to the normal disk-dominated type II migration determined solely by the viscosity of the disk, the migration in this regime is significantly slower and eventually even comes to a halt as the ratio continues to increase. Hence, the overall extent of the migration is significantly reduced and this despite the fact that higher mass stars are surrounded by higher mass disks, which, for a given planet mass, lead to more efficient migration in type I and type II disk dominated regimes.

Note finally that for $\alphaD = 2$ and $\mstar = 2.0 \msun$, a large number of warm super-Jupiter planets are predicted by the model (see Fig. \ref{a_m_alphaD2}), even when the disk lifetime is reduced compared to disks orbiting solar mass stars. These planet are not detected by present day RV surveys, and therefore, such a value of $\alphaD$ is excluded in the framework of our model.

\subsection{Detection rate and properties of observed planets}

When comparing the planet population that formed orbiting stars with masses 0.5 $\msun$, 1.0 $\msun$ and $2.0 \msun$ in Fig. \ref{a_m_nommodel}, \ref{a_m_alphaD0} and \ref{a_m_alphaD2},  we note that the number of detectable gas giant planets increases  with the stellar mass considered. In all cases, we computed the formation of planets around 30'000 stars and found respectively the following number of \red{observable} planets (with $K>10$ m/s and a period lower than 5 years) 500-600, 900 and 800-1600 depending upon the value of $\alphaD$.

We have compared the fraction of detectable planets for the different cases considered here, with the detection rate of planets orbiting different type of stars. Considering only planets more massive than five times Jupiter, and located between 0.5 and 2.5 AU from the central star, Lovis and Mayor (2007) concluded that the detection rate is 5 times higher around A stars (5 planets meeting the criterion), compared to FGK stars. In the case of M stars, no such planets have been detected. However, it should be noted that their sample included stars much more massive than 2.0 $\msun$. More recently, Bowler et al. (2009) have shown that the detection rate of similar planets around intermediate-mass stars is between 1.8 and 7 times higher than around solar type stars. 

We have compared the detection rate in our model for the 0.5, 1.0 $\msun$ and 2.0 $\msun$ cases by normalizing the number of detectable planets orbiting stars of a given mass to the number orbiting solar mass stars. In order to decide which planet is detectable, we used the same criterion as in Lovis and Mayor (2007). The ratio is found to range from 1 to 0.3 (for 0.5 $\msun$ stars) and from 1 to 3 (for 2.0 $\msun$ stars) for $\alphaD$ ranging from 0 to 2. Our ratio is therefore compatible with the two aforementioned observational results only if large values of $\alphaD$ are assumed. However, as pointed our in Sect. \ref{hotplanets}, such a large value of $\alphaD$ leads to large numbers of close-in massive planets orbiting 2.0 $\msun$ stars which are not observed. 

This apparent contradiction begs for an explanation. In fact, its origin could be three fold: Either the predicted masses are systematically too low, or the predicted semi-major axis too large, or the absolute number of planets predicted to form is too low. As we shall argue below, we believe the later explanation is correct and is a consequence of our assumption of a single embryo per system. 

\begin{figure}
\resizebox{\hsize}{!}{
\includegraphics[angle=0]{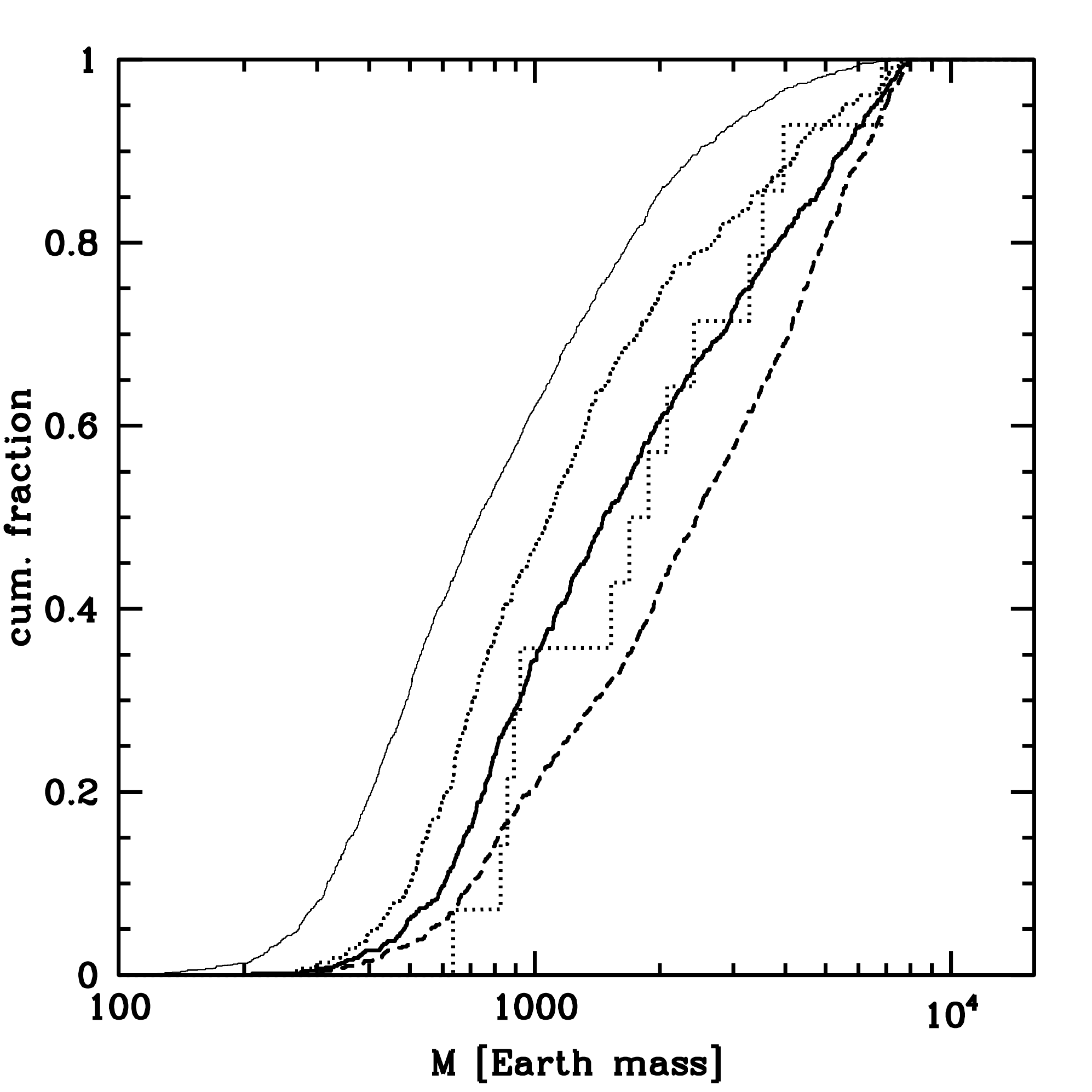}
\includegraphics[angle=0]{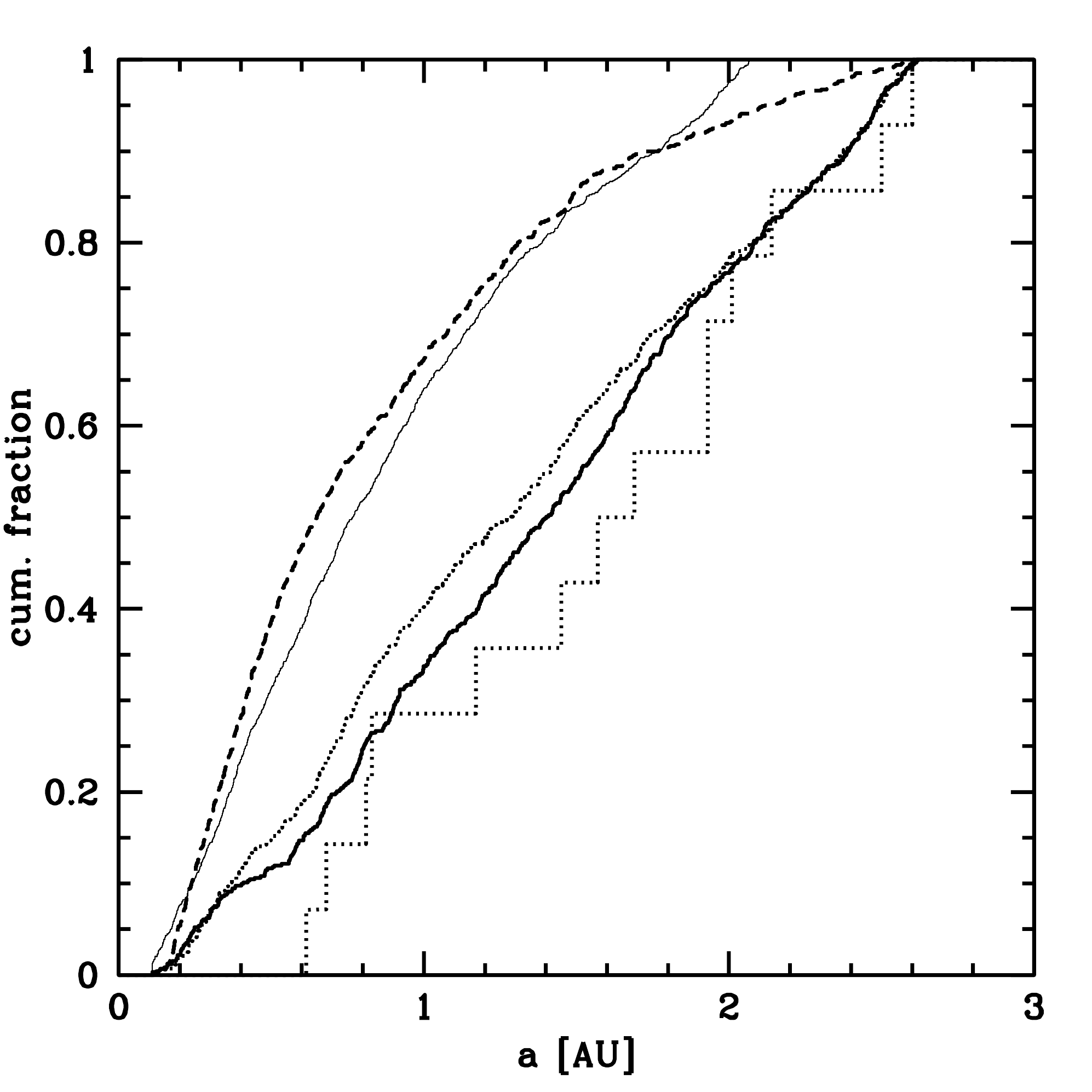}}
\caption {Cumulative histograms of minimum mass (left) and semi-major axis (right) of synthetic and observable planets orbiting high mass stars. In each case, we only take into account planets  whose induced Doppler semi-amplitude is larger than 30 m/s, and with periods lower than 3 years. Moreover, we exclude planets located at distances lower than 0.1 AU (the internal limit of our computational disk). For synthetic planets (heavy lines), we use the formation model around 2.0 $\msun$ stars, whereas we have selected planets discovered around stars with mass between $1.8 \msun$ and $3.0 \msun$ (step-like lines).  The three heavy lines are for the formation model around a $2.0 \msun$ star, for $\alphaD$ equal to 1.2 (solid line), 0 (dotted line) and 2 (dashed line).  The thin solid line is for a primary mass of 1.0 $\msun$. The results of statistical comparison between the different curves is indicated in Table \ref{tableKS_2.0}.} 
\label{cumul2.0_lowmass}
\end{figure}

\begin{figure}
\resizebox{\hsize}{!}{
\includegraphics[angle=0]{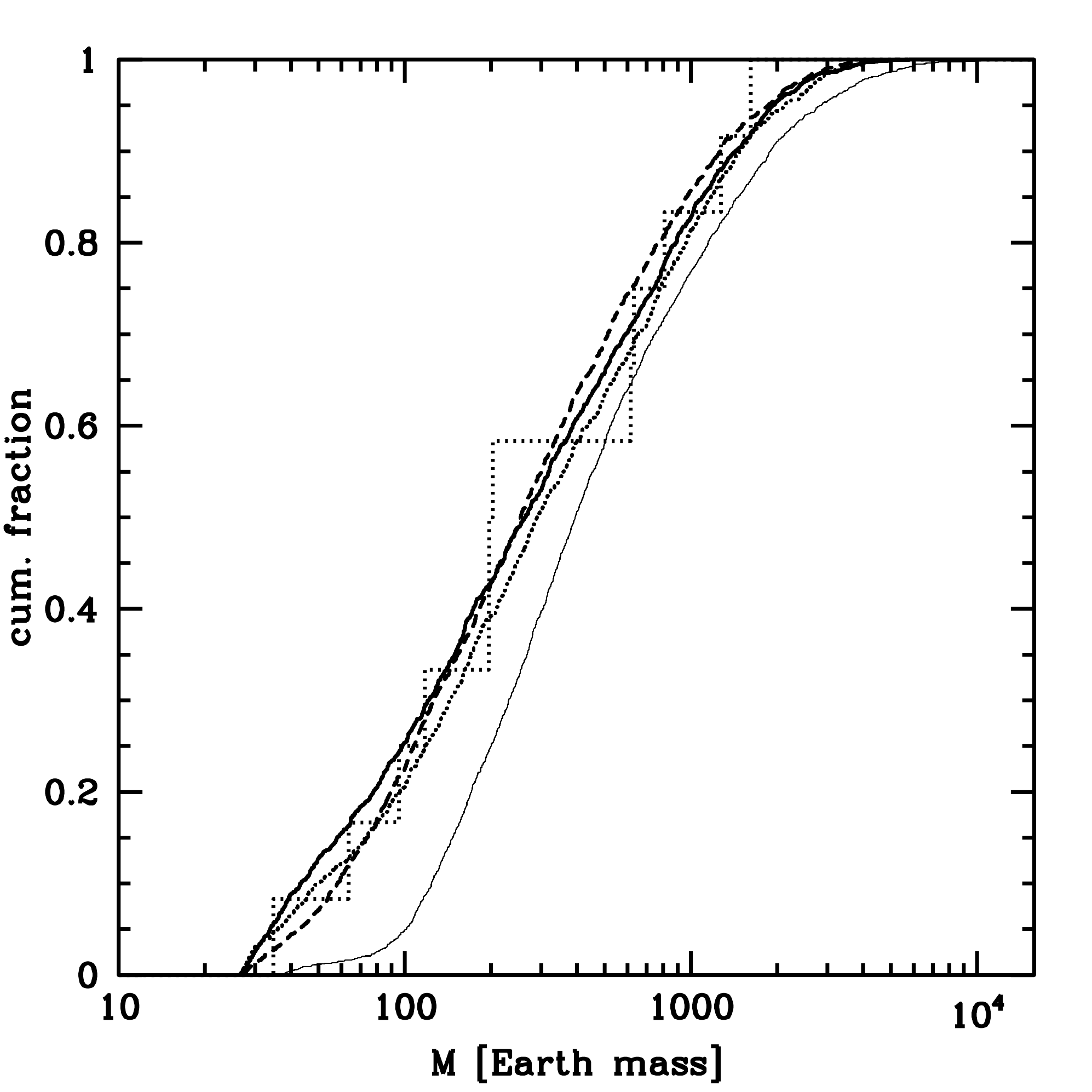}
\includegraphics[angle=0]{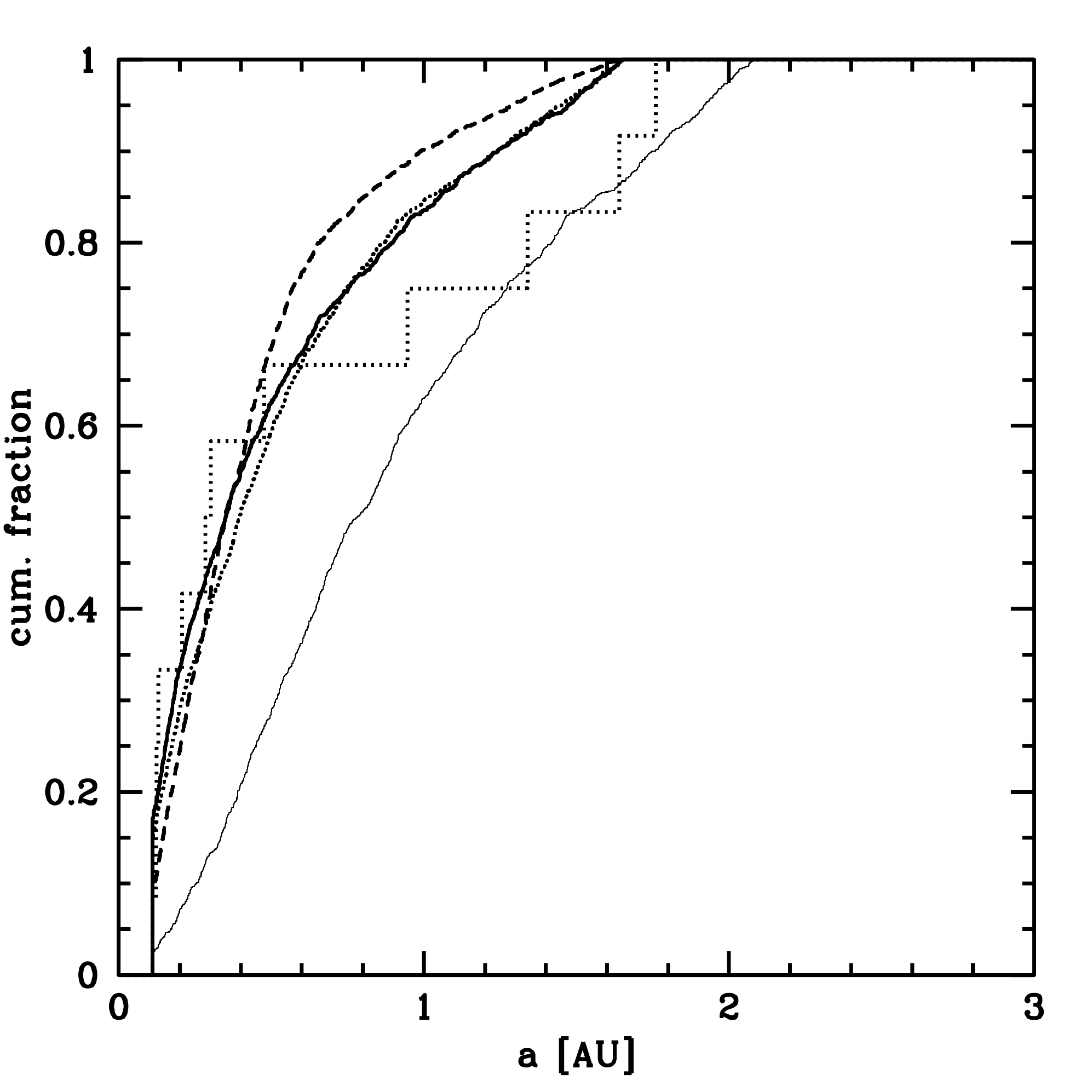}}
\caption {Cumulative histograms of minimum mass (left) and semi-major axis (right) of synthetic and observable planets around M stars. In each case, we only take into account planets  whose induced Doppler semi-amplitude is larger than 10 m/s, and with periods lower than 3 years. Moreover, we exclude planets located at distances lower than 0.1 AU (the internal limit of our computational disk). For synthetic planets (heavy lines), we use the formation model around 0.5 $\msun$ stars, whereas we have selected planets discovered around stars with mass between $0.3 \msun$ and $0.8 \msun$ (step-like lines). The three heavy lines are for the formation model around a $0.5 \msun$ star, for $\alphaD$ equal to 1.2 (solid line), 0 (dotted line) and 2 (dashed line). The thin solid line is for a primary mass of 1.0 $\msun$. The results of statistical comparison between the different curves is indicated in Table \ref{tableKS_0.5}.} 
\label{cumul0.5}
\end{figure}

We begin by looking at the cumulative histograms of minimum masses and semi-major axis of the observed and synthetic planets orbiting stars of 0.5 and 2.0 $\msun$ (Fig. \ref{cumul2.0_lowmass} and \ref{cumul0.5}). In these diagrams, we only considered planets  with orbital period shorter than three years and inducing a Doppler semi-amplitude larger than a fixed value (see caption)\footnote{Moreover, we note that only planets less massive than $20 \mjup$ at two sigmas are included in the Extrasolar planet encyclopedia (see http://exoplanet.eu/README.html). Therefore, it is difficult to compare models and observations in the very massive planets domain. We therefore exclude planets more massive than 25 $\mjup$ (to account in an approximate way for the 2 sigmas possible uncertainty in planetary masses) from the models and observations.}. The actually detected planets meeting these criteria were taken from the Extrasolar planet encyclopaedia (http://exoplanet.eu/) as of December 1, 2009. Finally, we plot on the same figures the distribution of the minimum masses and semi-major axis of the synthetic planets corresponding to the nominal model ($\alphaD=1.2$) assuming a 1.0 $\msun$ star. 

From these figures as well as from the statistical tests presented in Table \ref{tableKS_2.0} and Table \ref{tableKS_0.5}, we first note that the nominal model calculated for solar type stars fails at reproducing the characteristics of the detected planets orbiting either M or A stars even though it was successful at reproducing the properties of planets orbiting solar type stars (see paper II). 

As far as the planet population orbiting $2.0 \msun$ stars are concerned, the masses and semi-major axis distributions obtained with the nominal model ($\alphaD=1.2$) are in reasonable agreement with observations. The main differences being that the models predicts slightly smaller semi-major axis and a larger number of massive planets (see Fig. \ref{cumul2.0_lowmass} and Table \ref{tableKS_2.0}).

\begin{table} 
\begin{center}
\caption{Results of Kolmogorov Smirnov tests comparing theoretical and observed curves presented in Fig. \ref{cumul2.0_lowmass}, for the
distribution of planetary mass ($\mplanet$) and planetary semi-major axis ($\aplanet$).}
\begin{tabular}{llll}
\hline
& & $\mplanet$  & $\aplanet$  \\
\hline
$\mstar$ & $\alphaD$ & KS & KS \\
\hline
2.0 & 1.2 & 67 \% & 76 \% \\
2.0 & 0 & 10 \% & 49 \% \\
2.0 & 2 & 36 \% & 0.2 \% \\
1.0 &  & 0.2 \% & 0.3 \% \\
\end{tabular}
\label{tableKS_2.0}
\end{center}
\end{table}

In the case of planets orbiting M stars, the minimum mass is well reproduced for all values of $\alphaD$ considered while the semi-major axis distribution is not as well reproduced in particular for the $\alphaD=2$ case. The main difference is an underestimation of the semi-major axis of planets located at distances larger than about 0.5 AU. However, note that comparing the semi-major axis distribution of synthetic planet orbiting 1.0 $\msun$ stars with the distribution of actually detected planets orbiting M stars yields results (see Table \ref{tableKS_0.5}) statistically considerably less significant thus showing again the importance of the role played by the mass of the central star. 

The fact that the properties of detectable synthetic planets orbiting both 0.5 and 2.0 $\msun$ stars are similar to the observed ones lead us to believe that the discrepancy in the detection rate mentioned above is due to the fact that the number of planetary embryos increases with the mass of the central star. From a purely logical point of view this has to be expected for two reasons. First, due to the scaling of the mass of the disk with the mass of the primary, more mass is available at a given metallicity to form embryos. Second, since these embryos grow faster in orbit of more massive stars, they are also more likely to become giant planets that will eventually be detected. 

\begin{table} 
\begin{center}
\caption{Results of Kolmogorov Smirnov tests comparing theoretical and observed curves presented in Fig. \ref{cumul0.5}.}
\begin{tabular}{llll}
\hline
& & $\mplanet$ & $\aplanet$ \\
\hline
$\mstar$ & $\alphaD$ & KS & KS \\
\hline
0.5 & 1.2 & 92 \% & 83 \% \\
0.5 & 0 & 74 \% & 78 \% \\
0.5 & 2 &  87\% & 55 \% \\
1.0 &  & 11 \% & 1 \% \\
\end{tabular}
\label{tableKS_0.5}
\end{center}
\end{table}

Finally, we point out that for the planet populations orbiting both 2.0 $\msun$ and 0.5 $\msun$ stars, the nominal value for $\alphaD$ provides systematically the best fit to observations.  Interestingly enough, we recall that we determined the $\alphaD = 1.2$ value not by fitting characteristics of planet population but by comparing accretion rates obtained by our alpha-disk model with observational determinations of this rate. Hence, amongst the values tested, the the nominal value of $\alphaD$ is the value that consistently reproduce best both disk and planet observations.

\subsection{Metallicity effect}
\label{effetZ}

In our formation model, the metallicity of the central star enters only through the value of the planetesimals-to-gas ratio $\fpg$. The intrinsic \red{difficulties} in relating [Fe/H] and $\fpg$ are discussed in Paper I, and we adopt here the same prescription, namely [Fe/H]$=\log (\fpg / \fpgsun)$, with $\fpgsun = 0.04$ regardless of the mass of the primary (see Paper I for the justification of the $\fpgsun$ adopted value). 

\begin{figure*}
\includegraphics[angle=0,width=18cm]{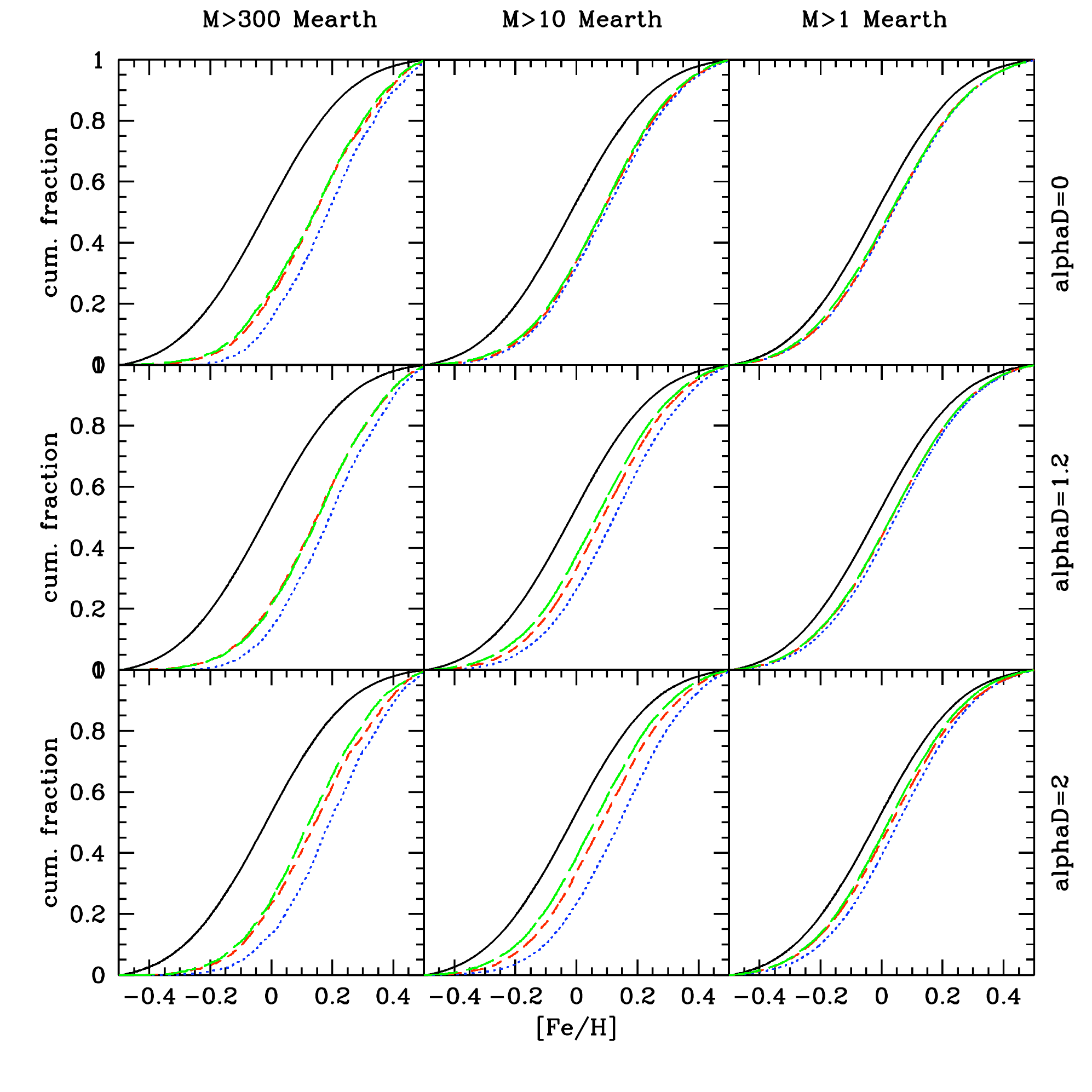}
\caption {Cumulative histograms of metallicities of all stars and of stars \red{harbouring} a planet more massive than a given threshold (see top of the figure). The solid black line is the cumulative histogram of all stars considered in this work, which does not depend on the mass of the central star. In each panel, the blue dotted curve is the cumulative histogram of 0.5 $\msun$ stars with a massive planet, the short dashed red line is for 1.0 $\msun$ stars, and the long dashed green curve is for 2.0 $\msun$ stars.} 
\label{effetZmass}
\end{figure*}

Fig. \ref{effetZmass} shows the \red{cumulative} metallicity distributions of stars without planets and stars with a planet more massive than a given mass. To show the effect of varying the mass of the central stars, these distributions have been computed for stars with masses 0.5 $\msun$, 1.0 $\msun$, and 2.0 $\msun$ as well as for the three different values of $\alphaD = 2$. For reference, the metallicity distribution for all stars in the sample, regardless if they are hosting planets or not, is also drawn on the same figure. 

The overall effect of metallicity on the formation of planets can be measured by the shift between the different colored curves (stars with planets) and the reference curve (all stars). The shift between the colored curves is a measure of the differential metallicity effect between stars of different masses but all having planets. We note that in all cases this differential metallicity effect is much smaller than the overall effect. Hence, we conclude from our models that the metallicity effect is only weakly depended upon stellar mass. 

From the figure, it is clear that a metallicity effect is present in all cases in the sense that stars hosting planets are, on average, more metal rich (colored lines are shifted with respect to the right of the black line). Comparing the three rows of Fig. \ref{effetZmass},  we also see that the metallicity effect also depends on the mass of the disk itself. Indeed, in the case of $\alphaD =0$ the effect is very similar for the three stellar masses considered. On the other hand, for $\alphaD =1.2$ and, even more importantly for $\alphaD = 2$, the effect increases as the mass of the primary decreases. This can be easily understood by recalling that the formation of a planet is, to first order,  dependent upon the total amount of mass in form of planetesimals.  In our model, this mass depends upon the product of the gas surface density (or the gas disk mass) and of the dust-to-gas ratio, which itself is a growing function of the metallicity. 

When the disk mass scales with stellar mass ($\alphaD = 1.2$ and $\alphaD= 2$), the disks orbiting 2.0 $\msun$ stars are much more massive than those orbiting 0.5 $\msun$ stars. Therefore, the minimum metallicity required to form a massive planet is correspondingly lower for massive stars than lower mass stars. Conversely, for $\alphaD = 0$, the metallicity effect is much less dependent upon stellar mass since for all masses, the total amount of solids available is only varying with metallicity. 

\begin{figure*}
\includegraphics[angle=0,width=18cm]{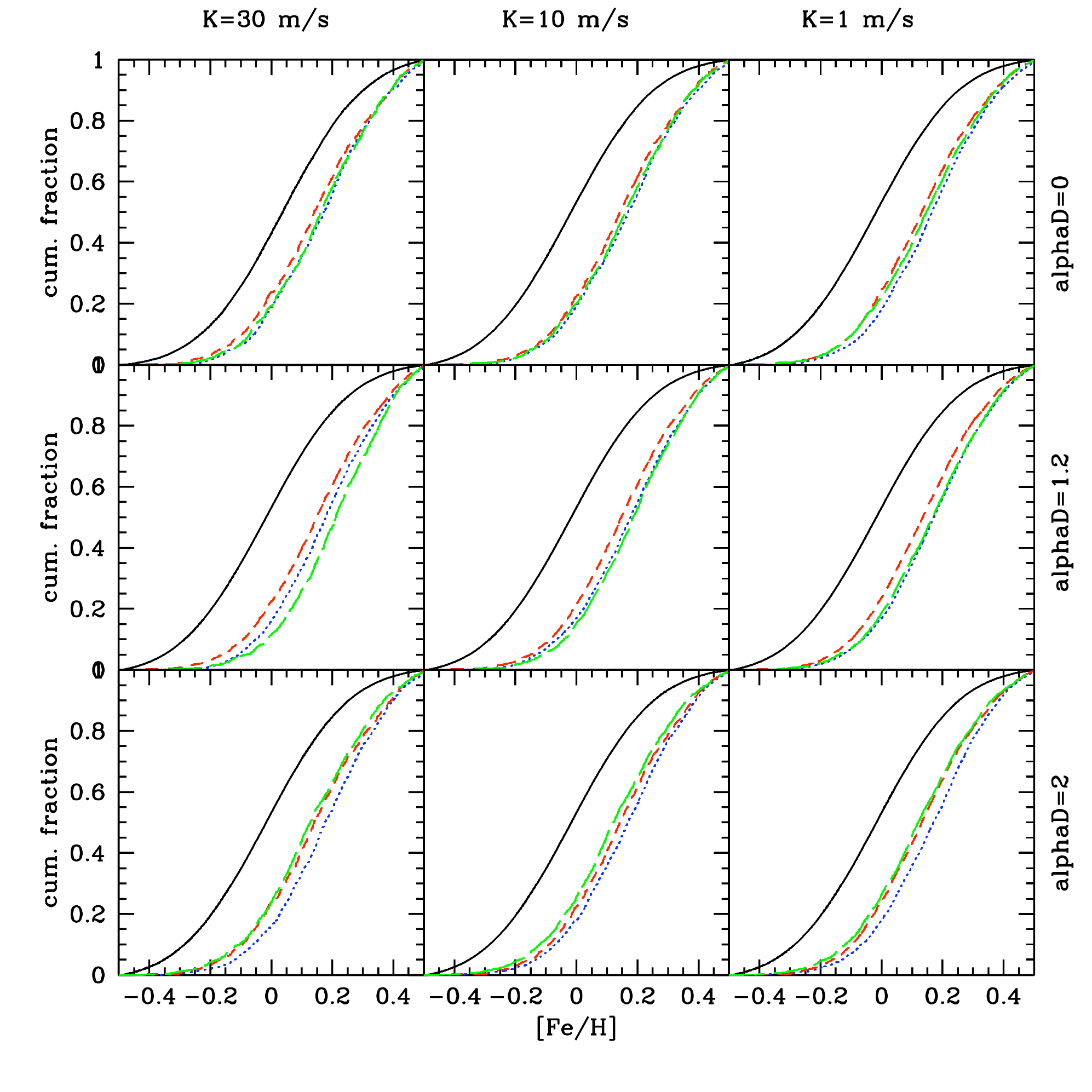}
\caption {Cumulative histograms of metallicities of all stars and of stars with an observable planets. 
The solid black line is the cumulative histogram of all stars considered in this work, which does not depend on the 
mass of the central star. In each panel, the blue dotted curve is the cumulative histogram of 0.5 $\msun$ stars with an observed planet, the short dashed red line is for 1.0 $\msun$ stars, and the long dashed green curve is for 2.0 $\msun$ stars. A planet is observable if it produces a RV amplitude larger than a given K, and has a period lower than 5 years. The different assumed values of K and $\alphaD$ are indicate on the top and right of the figure. Only planets more massive than the Earth are considered in the figure.} 
\label{effetZall}
\end{figure*}

Observationally, the metallicity effect is not defined by the existence of a planet more massive than a given value, but by the presence of an observable planet, with and induced \red{D}oppler shift larger than a given value. Fig. \ref{effetZall} present curves similar to Fig. \ref{effetZmass}, except that we now consider planets  with periods less than five years but inducing a \red{D}oppler shift higher than 30, 10 or 1 m/s. In this case, both the mass and the semi-major axis of the planets are important as they enter both in determining the amplitude of the \red{Doppler} shift.  

For the two first cases, $\alphaD = 0$ and $\alphaD = 1.2$, the effect of metallicty on the distribution is somewhat higher for 2.0 $\msun$ stars. This is a consequence of the reduced migration resulting from shorter disk lifetimes already pointed out in Sect. \ref{hotplanets}. Since planets orbiting massive stars are formed, on average, at larger distances from their parent star, they induce a given Doppler shift only if they are massive enough. As seen above, larger planets are formed provided the total amount of solids is large enough. This condition is easier to fulfill in the case of a strong scaling of the disk mass with the stellar mass. It is therefore no surprise that the minimum metallicity required to form a massive planet (a planet inducing a large stellar reflex motion) is somewhat lower for larger mass stars when $\alphaD = 2$.

\subsection{Composition}
\label{section_compo}

\begin{figure*}
\includegraphics[angle=0,width=18cm]{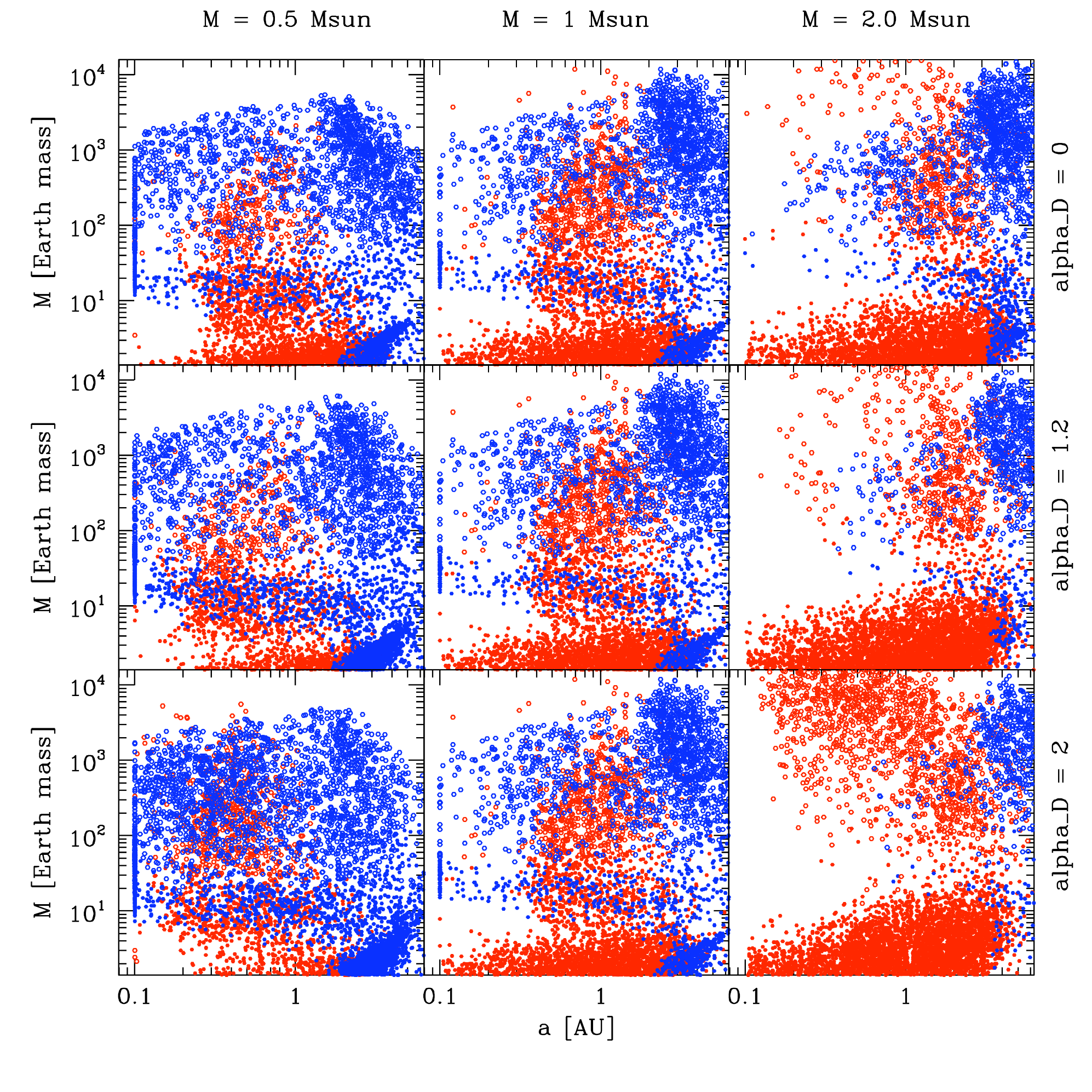}
\caption {Composition of planets for different values of the primary mass and of $\alphaD$. Red symbols are planets that have not accreted icy planetesimals, whereas blue \red{symbols} are for planet which have accreted a fraction of their solid mass in the form of icy planetesimals. Filled symbols are for planets whose \red{envelope}  mass if lower than heavy elements mass, whereas open symbols are for planets whose envelope mass is larger.} 
\label{compo}
\end{figure*}

The temperature in our proto-planetary disk model is computed as a result of the heat dissipation by viscosity, and therefore depends upon the gas surface density. As a consequence, the location of the iceline depends on the initial disk mass, as well as on the mass of the primary (viscous dissipation scales with the Keplerian frequency), as indicated in Eq. \ref{eq_iceline}. We have compared the resulting composition of synthetic planets, for different values of the $\alphaD$ parameter, and for the three values of the mass of the central star considered so far. The results are presented in Fig. \ref{compo}.

In order to differentiate between the bulk composition of different planets, we have used different types of colors and symbols. Blue symbols represent planets that have accreted some icy planetesimals while red symbols are for planet having accreted only rocky planetesimals. The open symbols are for planets where the mass of accreted gas is larger than the mass of accreted solids (envelope dominated) while the solid symbols are for planets that have accreted more solids than gas (core dominated). We note that this bulk composition corresponds to the time at which we end the calculation namely when the gaseous disk disappears. Later events that could modify this composition are not taken into account. Such events could be, for  example, evaporation that could reduce the planetary gas mass envelope (e.g. Baraffe et al. 2004), or later accretion of icy bodies (see e.g. Morbidelli et al., 2000 in the case of the Earth).

From the figure, it is evident that the composition of our synthetic planets depends strongly on the mass of the primary star and, to a lower extent, to the value of $\alphaD$ and hence the disk mass (the correlation between the disk properties and the planetary composition is studied in more details in Mordasini et al., 2010). Planets located closer than a few AU from their central star are found to be, on average, dryer when they orbit more massive stars.  This is a consequence of the position of the iceline which varies in these different environments. Indeed, the iceline is located 1 to 2 AU further out when increasing the stellar mass from 0.5 $\msun$ to 2.0 $\msun$. This effect is further amplified for $\alphaD > 0$, when the mean disk mass increases as a function of the primary mass.

\section{Discussion and conclusions}

We have extended the population synthesis calculations presented in Mordasini et al. (2009a,b) to the formation of planets
orbiting stars of different masses. Since the focus of this work was to pinpoint the effect of changing the mass of the central star and the corresponding changes in disk properties, we used the same model than presented in these two papers. In addition, we have explored the consequences on the planet population of difference scaling between the mass of the primary and the mass of the proto-planetary disk as well as between different relations between the lifetime of the disk and the central stellar mass.

Our calculations show that the main effect of varying the central body's mass occurs through the variations in the disk properties. The change in the later determine the changes in the synthetic planet characteristics. Assuming disk lifetimes independent of stellar mass results in the presence of many hot planets orbiting 2.0 $\msun$ stars, which is not consistent with observations. In addition, models assuming a very strong scaling of the disk mass with the primary mass ($\alphaD = 2$) predict a lot of warm very massive planets orbiting 2.0 $\msun$ stars (see Fig. \ref{a_m_alphaD2}, right panels). This again seems to be excluded from present day planet observations. As a consequence, we exclude from our model framework such high values of $\alphaD$. On the other hand, if such high values were obtained from future disk observations, it would indicate a that our models still lack some other physical mechanism.

The dependance of the detection rate as a function of the primary mass is lower in our model as compared to observationally inferred (Lovis \& Mayor, 2007, Bowler et al. 2009). However, the predicted distributions of the planetary masses and semi-major axis around $0.5 \msun$ and $2.0 \msun$ stars are statistically comparable with the one observed around M dwarfs and A stars. We therefore conclude that the variation in the detection rate as a function of the central star mass are likely due to the fact that the number of planetary embryos, some of which leading to observable planets, is a growing function of the primary mass. This in turn
is likely to result from the larger amount of planetesimals available to form planetary embryos around massive stars.

The first important result demonstrated in this paper is therefore that the planetary population depends on the mass of the central star: indeed it is not possible to statistically reproduce properties of extrasolar planets
discovered around low and high mass stars, using planet formation models developed for Solar type stars, as presented in Mordasini et al. (2009a,b).
The second important result is that this dependance of planetary properties on stellar mass is the result of changes in protoplanetary disk properties for different types of stars. This shows the importance of
both future observational programs aiming at studying disk structure and properties, and the necessity to improve the description of protoplanetary disks in planet formation models. In particular, future papers will study 
the influence of irradiation by the central star (see Fouchet et al. 2010), and the presence of dead zones in these disks. 

\begin{figure*}
\includegraphics[angle=0,width=18cm]{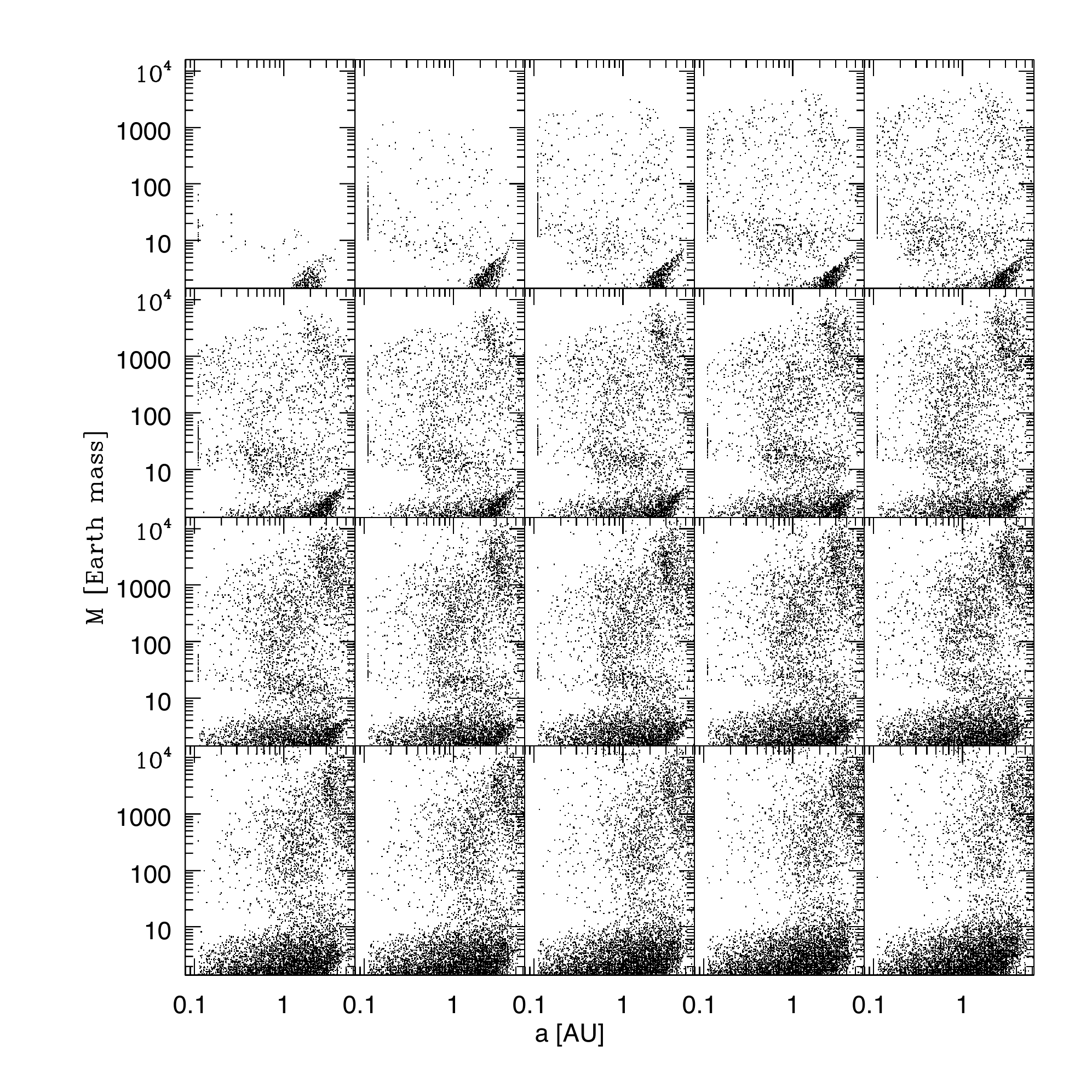}
\caption {Mass \textit{versus} semi-major axis for stars between 0.1 $\msun$ and 2.0 $\msun$. The $\alphaD$ parameter is equal to 1.2,
and disk lifetime are reduced for stars more massive than 1.5 $\msun$. 30000 stars are considered in each panel. First line, from left to right, masses
between 0.1 $\msun$ and 0.5 $\msun$, second line, from left to right, masses between 0.6 $\msun$ and 1.0 $\msun$,
third line, from left to right, masses between 1.1 $\msun$ and 1.5 $\msun$
fourth line, from left to right, masses between 1.6 $\msun$ and 2.0 $\msun$.} 
\label{allmasses}
\end{figure*}

Many future projects aim at detecting planets using techniques different from radial velocity surveys. One can cite transit surveys (e.g. CoRoT and KEPLER), astrometry surveys (GAIA), microlensing, etc... A common point of all these surveys is that they target, or will target, stars of very different masses. This is for example the case for CoRoT and KEPLER which observe all stars in a given magnitude range and a given field. Comparing population synthesis models with data observed by these surveys is therefore intrinsically more complicated, since one needs to include also properties of the star-disk systems which are not yet well known.  The results of these calculations are presented in Fig. \ref{allmasses} and can be used to compare theoretical results with the results of an observational survey. Such comparisons are presented elsewhere in the case of CoRoT (see Alibert et al. 2009), and for microlensing surveys (Alibert et al. 2010). As new observations, using different techniques, will become available, they will allow to put strong constraints on planet formation models. This will be the subject of future work.

\acknowledgements

This work was supported by the Swiss National Science Foundation, the European Research Council under grant 239605, and the Alexander von Humboldt Foundation.

\end{document}